# AI is the Strategy:

# From Agentic AI to Autonomous Business Models onto Strategy in the Age of AI

**René Bohnsack* & Mickie de Wet**

*Corresponding author: r.bohnsack@ucp.pt

**Abstract**

This article develops the concept of Autonomous Business Models (ABMs) as a distinct managerial and strategic logic in the age of agentic AI. While most firms still operate within human-driven or AI-augmented models, we argue that we are now entering a phase where agentic AI (systems capable of initiating, coordinating, and adapting actions autonomously) can increasingly execute the core mechanisms of value creation, delivery, and capture. This shift reframes AI not as a tool to support strategy, but as the strategy itself. Using two illustrative cases, getswan.ai, an Israeli startup pursuing autonomy by design, and a hypothetical reconfiguration of Ryanair as an AI-driven incumbent, we depict the evolution from augmented to autonomous business models. We show how ABMs reshape competitive advantage through agentic execution, continuous adaptation, and the gradual offloading of human decision-making. This transition introduces new forms of competition between AI-led firms, which we term synthetic competition, where strategic interactions occur at rapid, machine-level speed and scale. It also challenges foundational assumptions in strategy, organizational design, and governance. By positioning agentic AI as the central actor in business model execution, the article invites us to rethink strategic management in an era where firms increasingly run themselves.

## 1. Introduction

In March 2025, Amos Bar-Joseph, the CEO of Swan AI (getswan.ai), shared an extraordinary announcement: his startup of three founders aims to reach $30 million in annual recurring revenue without hiring a single additional human employee. Instead, Swan AI plans to scale by deploying an internal network of over twenty AI agents, effectively building an "intelligence network" rather than a traditional organization. In Bar-Joseph's words, the startup has "flipped the model on its head," replacing traditional departments and headcount growth with autonomous AI agents operating at enterprise scale.

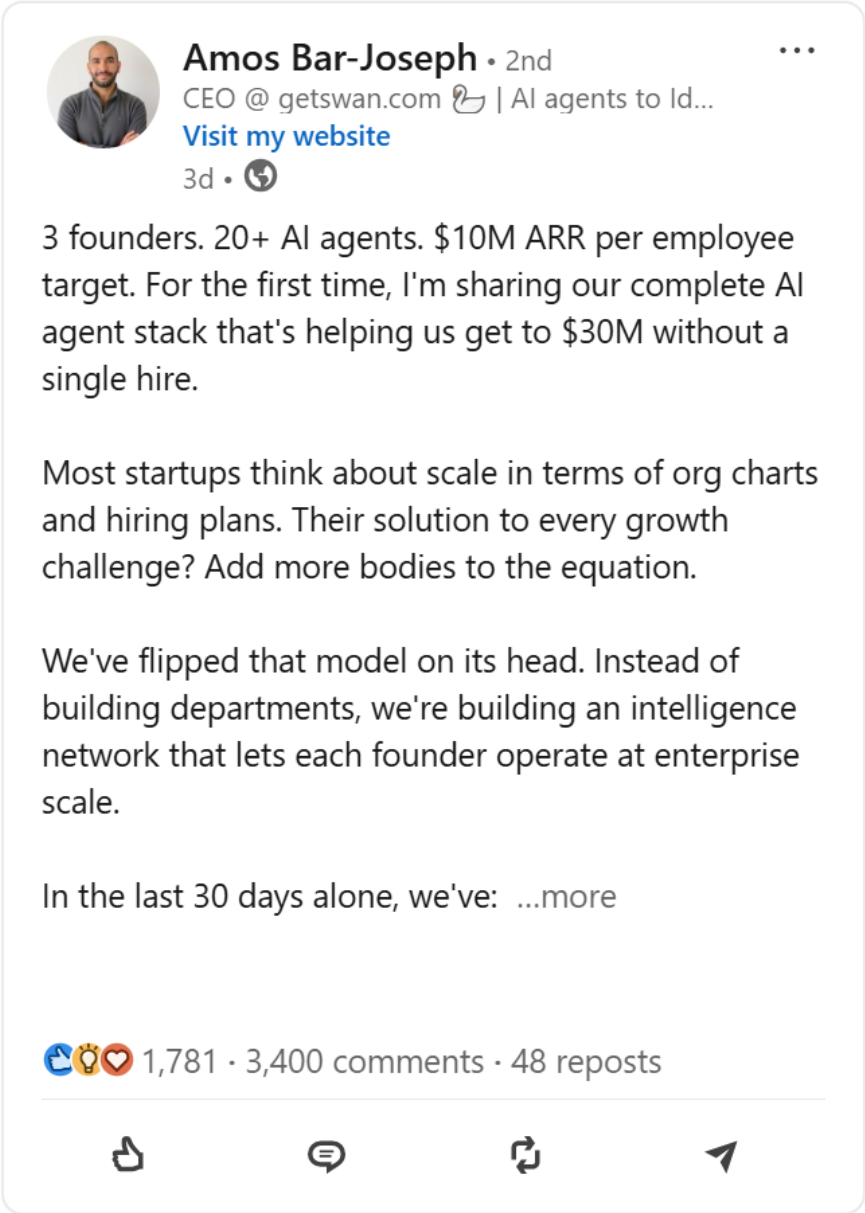

Figure 1: **Statement of the CEO of getswan.ai**

Only weeks later, Canadian-based Shopify CEO Tobi Lütke confirmed a strategic shift at one of the world's largest e-commerce platforms: AI adoption would become a baseline expectation at Shopify, not just for improving workflows, but as a first alternative to any new human hire. In a memo that he published publicly (being afraid it would be leaked regardless), Lütke emphasized that the default assumption across Shopify was now to ask *"Can this job be done by AI?"* before opening a new human position.

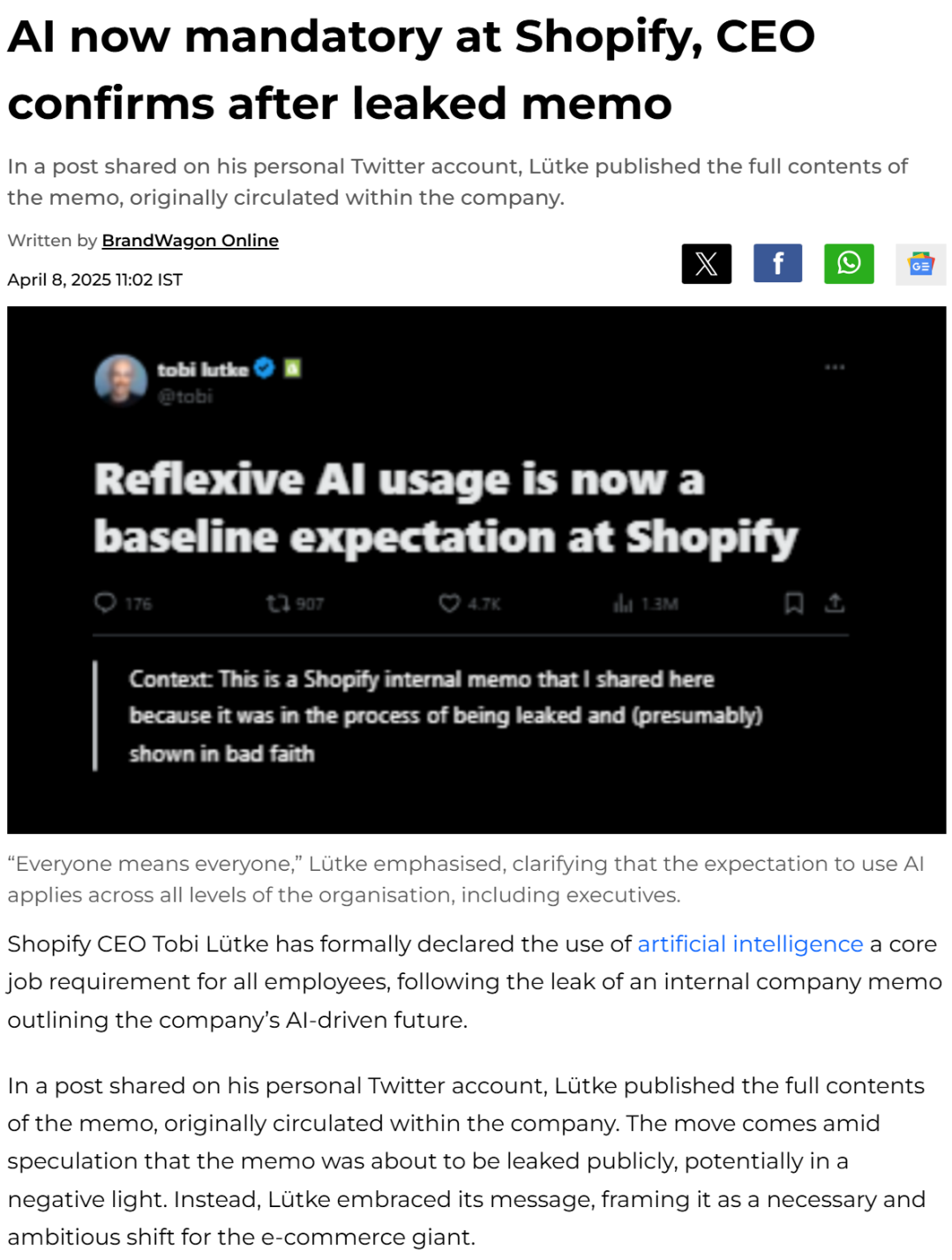

Figure 2: **Report on memo of Shopify CEO**

These announcements, although different in context—a hyper-growth startup and a tech incumbent—point to a profound shift. Companies are moving beyond simply integrating *generative AI* to assist employees with tasks like drafting emails, creating marketing material, or helping with analyses. They are beginning to integrate *agentic AI* —AI systems that can autonomously initiate, coordinate, and complete complex tasks with very limited or no human oversight.

We define *agentic AI* as a specific subset of generative AI distinguished by its autonomy: rather than merely generating content upon request, agentic AI systems take initiative, manage workflows, interact with other systems or agents, and continuously learn and adapt based on goals (Wiesinger et al. 2024). While generative AI generates outputs, *agentic AI generates and executes actions*, making it the crucial enabler for firms transitioning from human-augmented workflows toward *autonomous business models*.

In this article, we propose that the integration of agentic AI fundamentally alters the link between *strategy*, *business models*, and *tactics* (cf. Casadesus-Masanell and Ricart 2010). In earlier phases, AI served on a tactical level - supporting human workers within existing business models. However, as firms adopt agentic AI, the possibility emerges that AI systems themselves configure and operate significant parts of the business model autonomously. Over time, this leads to the emergence of *Autonomous Business Models (ABMs)* —systems in which AI agents handle core value creation and delivery mechanisms with minimal human intervention.

To explore this trajectory, we first present the real-world case of Swan AI (getswan.ai) as an early example of an agentic AI-first business model. We then introduce a *hypothetical extension* using Ryanair, imagining how an incumbent could evolve toward an autonomous model through gradual AI integration. These vignettes highlight different routes for startups and incumbents.

Finally, we develop a framework explaining the progression from human-augmented business models to fully autonomous ones, and discuss the implications for competitive dynamics, managerial roles, and future research. We argue that the rise of agentic AI leads not just to more efficient firms, but eventually to *synthetic competition* —a new form of competition where autonomous AI agents operate at ultra-fast speeds, learning and competing without direct human control.

By focusing explicitly on agentic AI as the catalyst, we aim to contribute to the emerging conversation on how AI transforms business model theory and strategic management, not just as a tool, but as an autonomous actor within the firm.

## 2. Agentic AI: Foundations and Applications

The announcements by Swan AI and Shopify introduced in the previous section signify not just substantial advancements in AI's augmentative role in the workplace, but a transformative shift towards the employment of AI systems capable of completely autonomous action. To understand the profound implications for strategic management and business models, it is essential to clearly show what distinguishes these advanced systems from earlier forms of AI. *Agentic AI* refers to artificial intelligence algorithms capable of autonomous decision-making and executing certain actions to achieve predefined goals with minimal ongoing human oversight. Unlike earlier forms of generative AI, which respond passively to prompts by producing content (such as text or images), agentic AI proactively initiates, coordinates, and completes complex tasks. At its core, agentic AI combines the linguistic and reasoning capabilities of Large Language Models (LLMs) with strategic decision-making enabled by reinforcement learning and planning algorithms (Rai et al. 2019; Wiesinger at al. 2024). The result is a system that can autonomously set intermediate objectives, adapt to changing conditions, and learn from outcomes, thus demonstrating genuine agency rather than mere responsiveness.

While generative AI tools like ChatGPT create outputs in response to specific instructions, an agentic AI moves beyond generation to action. For instance, generative AI can draft marketing copy when prompted by a marketer, whereas an AI agent would be able to independently identify target customers, write tailored messages, execute inbound or outreach campaigns, analyze results, and continuously refine its approach based on feedback or leads generated, completely autonomously. Agentic AI therefore represents a critical evolutionary step in AI technology, transforming it from a supportive tool into an autonomous actor within the organizational framework. Agentic AI is still nascent, yet it has already demonstrated substantial potential across various core business areas (Xu et al. 2024). We illustrate this with critical applications, highlighting how agentic AI can fundamentally transform value creation, operational efficiency, and strategic execution.

In *Sales and Marketing* agentic AI can automate end-to-end sales processes, significantly enhancing scalability and productivity. For example, Swan AI's autonomous sales agents independently identify and qualify potential customers (aka determine their willingness to buy), execute outreach via email or other channels, follow up persistently, and schedule meetings (Swan AI n.d.). This approach allows firms to scale sales operations without proportional increases in human labor, creating substantial efficiencies and enabling autonomous, 'hockey stick' revenue growth.

In *Operations contexts*, agentic AI can streamline complex logistical decisions through continuous, real-time optimization. For instance, airline and logistics companies employ agentic systems that autonomously manage fleet maintenance schedules, dynamically adjust inventory levels, and optimize routing based on real-time data such as demand forecasts, and weather or traffic conditions. These autonomous operational agents not only enhance efficiency but also rapidly respond to changing environmental conditions without human intervention (Aylak 2025; Mohammed et al. 2025).

Finally, in *Customer Service* agentic AI has the potential to improve customer service capabilities by independently resolving complex customer interactions. Unlike traditional chatbots restricted to simple issues and scripted replies, advanced agentic customer service systems could autonomously interpret customer intent (Shah et al. 2023; Bodonhelyi et al. 2024), manage full-service processes like account adjustments or refunds and, when necessary, escalate complex scenarios to human operators. This autonomy could help improve service responsiveness, consistency, and customer satisfaction.

These practical applications underscore agentic AI's potential to enable entirely new operational logics. When critical business processes become autonomously driven by intelligent AI systems, it prompts organizations to rethink not just efficiency, but the very architecture of their business models. This profound shift lays the groundwork for the emergence of Autonomous Business Models (ABMs), where AI agents serve as central executors of value creation, delivery, and capture mechanisms. The following section provides a deeper conceptual exploration of these dynamics, clarifying the distinction between traditional strategies, business models, and tactics, thereby situating the implications of agentic AI within established strategic management frameworks.

## 3. Conceptual Background

### Strategy vs. Business Model vs. Tactics

Strategy scholars have long distinguished between the concepts of strategy, business models, and tactics to highlight different layers of decision-making within a firm. *Strategy* is generally seen as the high-level plan or vision that a firm adopts to achieve competitive advantage - essentially choosing a way to compete. In an oft-cited analogy, Casadesus-Masanell and Ricart (2010) suggest that if strategy is akin to designing and building a car, the business model is the car itself, and tactics are how one drives the car. The business model thus represents the logic of the firm's value creation and capture mechanisms: the architecture of how resources, activities, and offerings come together to deliver value to customers and convert that value into profits. *Tactics*, in turn, refer to the choices and actions a firm can take within the bounds of a given business model to improve its performance or respond to day-to-day challenges (for example pricing tweaks, marketing campaigns, or operational adjustments) without fundamentally changing the model.

In formal terms, a business model can be viewed as a system of interdependent activities and choices that determines how the firm creates and captures value (Amit and Zott 2001; Teece 2010). It encompasses elements such as the value proposition (what is offered and to whom), the revenue model (how the firm earns money), the resources and capabilities required, and the configuration of internal and external processes that deliver the offering. Strategy, by contrast, involves deciding which business model to employ (or whether to create a completely new one) in light of the competition and environmental conditions. As Casadesus-Masanell and Ricart (2010) argue, strategy is fundamentally about making choices (often under uncertain conditions and against rivals) that determine the business model a firm will operate. On the other hand, the business model is turning those strategic choices into a working structure. Tactics are the fine-tuning within that structure: given a certain business model (the "rules of the game" the firm is playing), tactical moves play within those rules without changing them.

This framework can be illustrated with a classic example: the airline industry. A full-service airline like British Airways historically pursued a differentiation strategy targeting business and premium leisure travelers, which translated into a business model featuring a hub-and-spoke network, multiple travel classes, extensive in-flight services, and travel-agent ticket distribution. Its tactics might include loyalty programs, fare sales in lean seasons, or marketing promotions - actions that refine performance but stay within the full-service model. In contrast, Ryanair adopted a low-cost strategy focused on price-sensitive travelers.

They chose a radically different business model: point-to-point flights, single-class no-frills service, secondary airports, direct online sales, and ancillary revenue streams, such as fees for extras. Ryanair's tactics—charging for priority boarding and baggage allowances, dynamically adjusting fares, and upselling add-ons—make sense only within the logic of its low-cost model. This example underscores how strategy shapes the choice of business model, which in turn defines what tactical moves are available or effective. Notably, a sound business model can be a source of competitive advantage in itself, especially if it is novel or hard to imitate (Zott and Amit 2010; Markides 2013). However, the business model must also align with the broader strategy and competitive environment. A misfit between strategy, model, or poor tactical execution can lead to failure even if each element is well-conceived in isolation.

Jacobides, Cennamo, and Gawer (2018) define a business ecosystem as a set of interconnected players (firms, and often other stakeholders) that collaborate and co-evolve capabilities around a shared value proposition, with each contributing a part of the solution. An ecosystem's ultimate aim is a joint value proposition that no single firm could achieve alone, for example the constellation of smartphone manufacturers, app developers, component suppliers, and telecom providers that collectively deliver the value we perceive as the "mobile experience." This perspective highlights that a firm's business model is not designed in isolation; it may deliberately depend on external complementors (as in Apple's iOS App Store model relying on third-party app innovation) or on participation in someone else's platform.

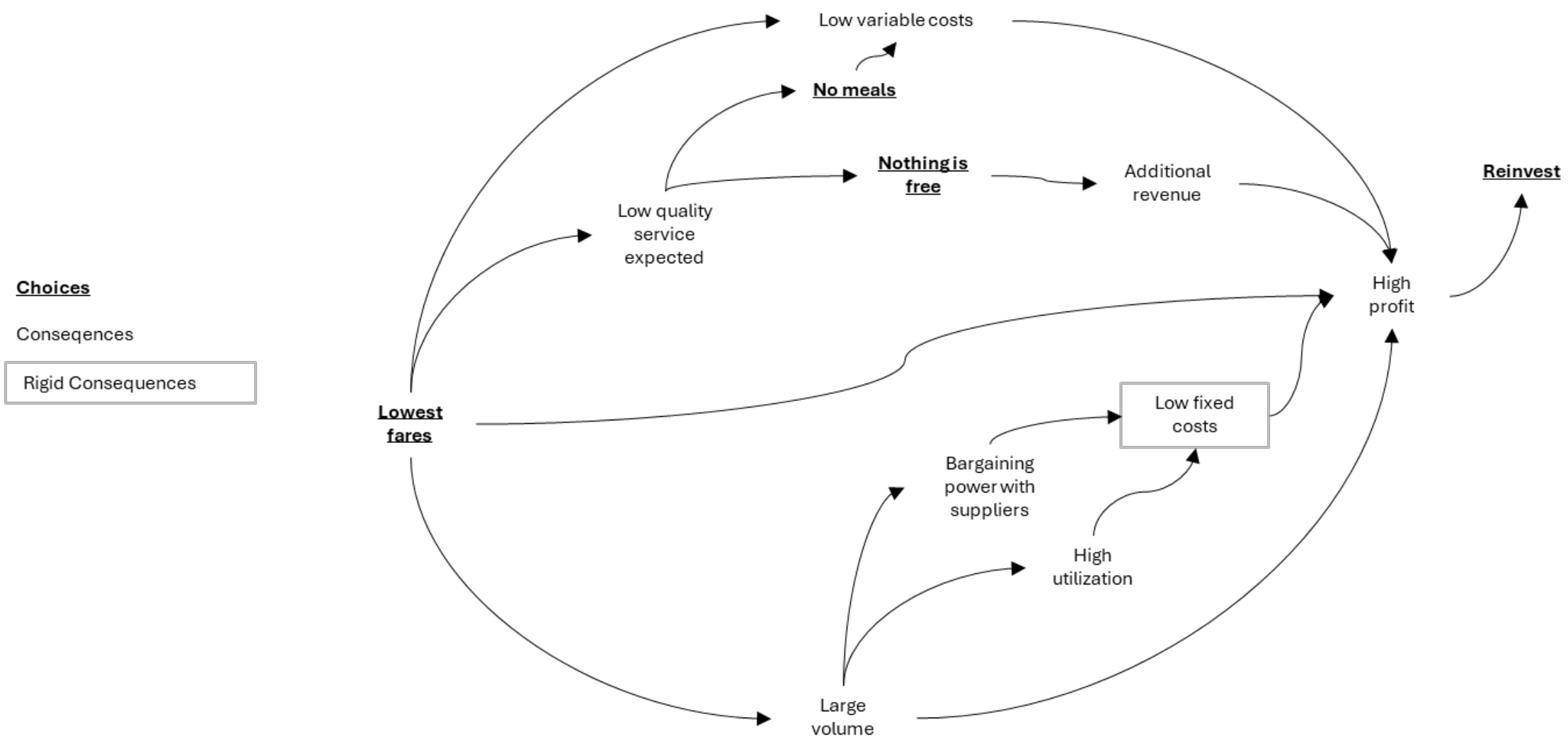

Figure 3: **Ryanair's business model (cf. Casadesus-Masanell and Ricart 2010), illustrating the link between Choices and Consequences**

## Business Models in an Evolving Digital Context

Over the past two decades, the business model concept has risen to prominence partly due to digital innovation as well as the emergence of new forms of value creation that traditional strategic analysis did not fully capture. Digital technologies enabled novel models, for example the platform business model, where a company like Uber or Airbnb creates a digital marketplace enabling transactions between producers and consumers, or the freemium model common in software, where basic services are free to attract a user base that can be monetized via premium upgrades or paid advertising. The literature has expanded to examine not just firm-level models but also ecosystems and industry architectures in which multiple firms' business models interact.

Indeed, a platform can be seen as one kind of business model that facilitates others — although, as Jacobides et al. (2021) clarify, a platform is not the same as an ecosystem. A platform is typically a technology or service that brings together different user groups under rules set by the platform owner (for example, Amazon's marketplace connecting sellers and buyers under Amazon's governance), whereas an ecosystem implies more distributed co-creation, co-ownership and co-evolution, without a single actor controlling all interactions.

These digital-era constructs matter for our discussion of AI and autonomous business models in two ways. First, they show how business models have become more fluid and boundary-spanning - today's competitive advantage often comes from network effects and complementarity (Jacobides et al. 2018; Iansiti & Lakhani 2020) rather than just internal efficiencies. Second, they underscore the increasing role of software and algorithms as core components of business models. In a platform model, for example, the platform's algorithms (for search ranking, matching, reputation scoring, dynamic pricing etc.) are central to how value is created and distributed among participants. In effect, many modern business models already embed algorithmic decision-making. Consider how Facebook's algorithmic news feed selection drives the user experience and engagement of the social media model, or how Uber's automated dispatch and pricing system coordinates supply and demand in real time. These are early examples of what we refer to as AI-augmented business models. In such cases, the firm's value creation logic crucially relies on AI or algorithmic components, although typically these operate under human-defined parameters and oversight. Management still sets objectives and policies and intervenes in exceptional cases, while AI handles well-defined tasks at scale (Rai et al. 2019).

**From AI Augmentation to Autonomy**

The current state of many firms can be described as AI-augmented: AI is used as a powerful tool within an otherwise traditional business model and organizational structure (Rai et al. 2019). The strategic rationale is often efficiency (cost reduction) or effectiveness (better decisions or personalization) in specific processes. For example, an e-commerce retailer might use machine learning to augment its product recommendation engine, increasing cross-sales, a manufacturer might deploy AI for predictive maintenance on equipment to reduce downtime, or a bank might even use AI to augment credit scoring, enabling faster loan decisions (Iansiti & Lakhani 2020). In each case, the business model (what the firm offers and how it makes money) might not fundamentally change; the retailer still sells products online for a margin, but AI improves how well the model performs.These augmentations can be significant. Amazon's recommendation algorithms reportedly generate a substantial portion of its sales by tailoring offerings to customers (Lopez 2025), illustrating how AI can augment value capture within an existing model. Yet, the governance of the business remains human-led given that AI provides inputs or automates sub-tasks, but humans typically make the overarching decisions such as merchandising strategy or pricing rules.

The frontier that firms are now approaching is one in which AI moves from a decision-support role to a decision-making and self-optimization role, giving rise to what we call Autonomous Business Models (ABMs).[1] In an ABM, algorithms do not only assist in executing the strategy —they effectively become the primary agents executing and adapting the business model. The notion is analogous to autonomous vehicles: just as a self-driving car can sense its environment and drive without a human behind the wheel, an autonomous business model would be capable of sensing market conditions, making operational decisions, and learning and adapting its tactics (and perhaps even elements of strategy) without constant human direction. This does not imply a complete absence of human oversight - as much as a self-driving car might still allow a human to set the destination and intervene in emergencies, a firm's owners and leaders would still set high-level goals and constraints. However, the day-to-day running of the business (responding to customer behaviors, optimizing pricing and inventory, customizing marketing, managing workflows) could be largely handled by an integrated system of agentic AI.

To make this concrete, we next present two illustrative vignettes: first, Swan AI (getswan.ai), representing a startup natively pursuing an autonomous model; and second, Ryanair, representing a hypothetical incumbent path toward increasing AI autonomy. These cases, while cursory, help to visualize what moving along the continuum from augmented to autonomous business models could look like in practice.

## 4. Vignettes: From AI-Augmented to Autonomous Business Models (ABMs)

**Swan AI: A Startup Built for Autonomy**

Swan AI (getswan.ai) is an illustrative example of a new venture aiming to disrupt a traditional service through autonomous agents. Founded in the early 2020s in Israel, Swan AI set out to "make sales human again" by automating the tedious early stages of the B2B sales process. The company offers businesses an AI-powered service that generates and nurtures sales leads autonomously. In essence, Swan AI's product is an AI sales agent – software that can identify potential customers, reach out with personalized emails or messages, follow up, and schedule meetings, performing much of the work a human sales development representative would typically do (Swan AI n.d.).

[1] See *Strategy Science* Call for Papers for the Special Issue on "Can AI Do Strategy? Exploring the Frontiers of AI-Augmented Strategic Decision Making," edited by Felipe Csaszar, Gwendolyn Lee, Peter Zemsky, and Todd Zenger.

Swan AI's business model is inherently digital and service-oriented: clients (other businesses) subscribe to its AI sales agent service, paying either a monthly fee or per-qualified-lead delivered. The value proposition is clear —more leads and less routine work, as the AI can engage thousands of prospects in parallel, something human sales representatives would struggle to do cost-effectively. What makes Swan AI interesting is the autonomy of its offering. Once the client sets the initial parameters (i.e. target customer profile, campaign goals, basic messaging templates), the AI agent takes over. It crawls the web and internal databases to compile lists of prospects matching the profile, drafts customized outreach messages (drawing on a large language model that can tailor tone and content to the industry and recipient), and manages the email or chat communication. It learns from responses, for example if certain subject lines or value propositions get better engagement, it will adapt its approach in future messages. When a prospect shows interest or wants to book a meeting, the AI agent can even interact with calendar systems to set up an appointment with a human salesperson from the client's team, at which point the "handoff" occurs and the human steps in to (hopefully) close the deal. In short, Swan AI's agent acts autonomously through the lead generation and qualification stages of the sales funnel, a role traditionally requiring significant human labor (Swan AI n.d.).

From a business model perspective, Swan AI represents an autonomous business model on a small scale. Internally, Swan AI itself likely operates with a tiny human team —the product (i.e. the AI sales agents) does most of the work that scales with the number of clients. The startup's cost structure will be dominated by computing resources and AI development, not by hiring large numbers of employees (Sjödin et al. 2021). Its revenue scales with how many clients and campaigns the AI can handle, which in turn depends on computing scale rather than linear hiring. This is a dramatic departure from a traditional sales outsourcing firm or telemarketing agency, which would have to hire more personnel as they take on more clients. Swan AI achieves scale without a proportional increase in headcount. Moreover, the more data Swan's AI agents gather on what approaches work in sales outreach (i.e. which phrases get responses, which industries have higher hit rates at certain times, and so forth), the smarter and more effective they become, creating potential data network effects (Gregory et al. 2022). Each new client's campaigns provide new training data and inputs that improve the AI for all clients (assuming the learning is aggregated). This means Swan AI's service could continuously improve and possibly gain a widening advantage if it captures a significant market share. This drives the concentration of market power and resembles the dynamics seen in other AI-rich platforms where "the more you use it, the better it gets."

Swan AI's model approaches autonomy because, for the specific domain of lead generation, the AI agents handle practically everything without human micromanagement. The startup's human team primarily focuses on developing the AI, maintaining quality (making sure the AI does not go off-brand or violate compliance rules in communications), and business development to acquire clients after the AI hands over warm, qualified leads. Over time, even some of those functions might be automated or augmented by AI (for instance, monitoring algorithms could flag when the AI agent needs new training data or when it encounters novel scenarios it is unsure about). Swan AI thus operates at the edge of an ABM: its core operational processes—the very thing clients pay for—are autonomous. The business model is almost "autonomous" in the sense that once the system is set up, value is created and delivered (qualified sales leads) with minimal ongoing human input. The AI essentially runs the service day-to-day.

Of course, Swan AI also highlights the current limits and challenges of ABMs. The scope of autonomy is relatively narrow (sales outreach), and there are guardrails: if the AI starts to perform poorly or encounters complex negotiations beyond its capability, human salespeople will step in for face-to-face discussions and the relationship building needed to close deals. Trust is another factor. Clients must trust Swan's AI to represent their company's brand professionally in communications, which Swan addresses by allowing initial customization and providing transparency or transcripts of AI interactions (Swan AI n.d.). This underscores that ABMs often require new forms of oversight and trust-building initiatives. Additionally, competition can emerge if the algorithms become commoditized —–if many companies can build similar AI agents, then having an autonomous model is not enough; how smart or proprietary your AI is, becomes the differentiator. We will discuss these implications later.

In summary, Swan AI exemplifies a startup leveraging AI to create a business model with high autonomy and scalability from the outset. Unlike an incumbent, it did not have to retrofit AI into an existing organization; it built the organization around the AI. This gives it agility and cost advantages but also exposes it to uncertainties of depending heavily on AI performance (Sjödin et al. 2021).

**Ryanair: An Incumbent Augmenting Its Business Model with AI**

Ryanair is Europe's largest low-cost airline and an exemplar of a tightly integrated, high-efficiency business model (cf. Casadesus-Masanell and Ricart 2010). Its strategy, since the 1990s, has been focused on cost leadership: offer the lowest fares by obsessively stripping out costs and monetizing ancillary services. This strategy translated into a business model with features such as a single aircraft type fleet (Boeing 737 for scale economies in maintenance and training), use of secondary airports with lower fees, no-frills service (i.e. no free meals or assigned seats originally), direct online sales (bypassing travel agents), and revenue from extras (e.g., fees for priority boarding, baggage allowances, and hotels and car rentals via partners). Notably, Ryanair was a pioneer in moving ticket sales online in the early 2000s, which was not only cost-saving but also gave the airline valuable data and control over pricing and customer interaction. Put differently, Ryanair's coherent set of choices formed a virtuous cycle: low base fares drove volume, which gave bargaining power over airports and suppliers, which lowered costs further, enabling even lower fares, and so on.

Against this backdrop, how has Ryanair incorporated AI, and does it move toward an autonomous model? Thus far, most evidence suggests that Ryanair uses AI and advanced analytics to augment its decision-making and operations in line with its existing model. For example, like all major airlines, Ryanair employs sophisticated yield management systems – algorithms that adjust ticket prices in real-time based on demand, booking pace, remaining capacity, and competitive pricing. These systems, rooted in operations research and now often enhanced with machine learning, automate a complex decision (setting the right price for each seat at each moment) far better than any human could. This contributes directly to Ryanair's value capture, optimizing revenue as much as possible from each flight while still undercutting competitors. Similarly, Ryanair likely uses AI for route and schedule optimization, simulating various network scenarios to decide which new city pairs to serve or how to adjust frequencies. Another area is predictive maintenance: AI models analyze sensor data from aircraft to predict component failures, allowing maintenance to be done proactively and minimizing costly aircraft downtime (Deslandes 2025). Adopting such AI-driven maintenance scheduling aligns perfectly with Ryanair's low-cost model (preventing delays and disruptions that would incur compensation costs or require reserve aircraft).

Customer-facing processes have also seen AI augmentation. Ryanair's mobile app and website utilize personalization algorithms to recommend add-ons (like hotel deals at the destination or travel insurance) based on customer history and context. Chatbot assistants handle common customer inquiries (for example, about baggage policies or flight status), reducing the need for call center staff (Amazon Web Services 2019). These applications of AI improve efficiency and customer experience within Ryanair's no-frills, digitally driven service approach. Importantly, however, the strategic logic remains very much human-driven —Ryanair's management decides where to expand, how to handle competitor moves (such as a price war or a new entrant on a route), and how to balance cost cuts with customer satisfaction. In other words, AI is used extensively but within boundaries set by managers. The business model has not fundamentally shifted; rather, it has been improved by AI. Ryanair's core value proposition (low fares) and its revenue model (ticket sales plus ancillary fees) are intact —AI just helps execute them more effectively (for example, more granular pricing, more targeted ancillary sales).

Could Ryanair push further toward autonomy? It is conceivable. In a future scenario, one could imagine Ryanair evolving toward a more autonomous configuration centered around an internal *AI Factory,* a concept adapted from Iansiti and Lakhani (2020). In such a model, Ryanair's AI Factory would serve as the operational heart of the airline, continuously ingesting data from booking trends, market prices, fleet telemetry, and customer behavior, dynamically orchestrating key business activities. Specifically, the AI Factory could autonomously handle *Fleet Management* (optimizing aircraft allocation, maintenance schedules, and rotations) and *Dynamic Pricing* (real-time fare adjustments based on predicted demand, competitor moves, and customer segmentation). Rather than merely assisting human planners, the AI Factory would *drive* operational decisions, learning and improving over time. Management's role would shift toward setting strategic guardrails and governance principles for the AI systems, ensuring alignment with regulatory requirements, brand positioning, and ethical standards (Iansiti and Lakhani 2020). We illustrate this envisioned evolution in Figure 4 below, conceptualizing the future Ryanair business model as a flywheel increasingly powered by its AI Factory.

*Mechanism 1: Agentic execution of core logic*

In an ABM, AI systems are no longer decision-support tools; they are the agents executing the

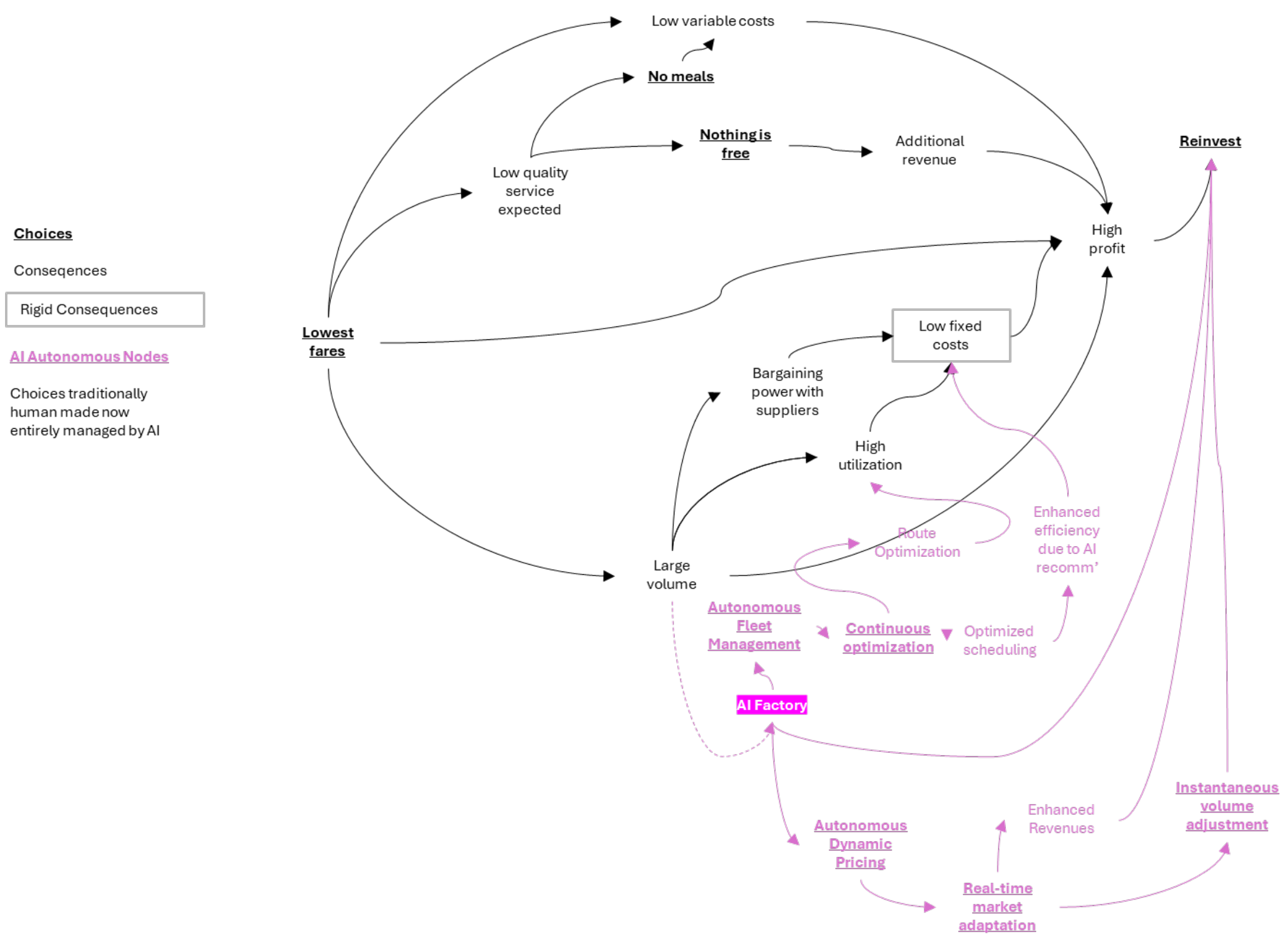

Figure 4: **Hypothetical Future Ryanair Business Model with AI Factory at the Core**

Together, Swan AI and Ryanair show a spectrum: from AI-augmented (Ryanair's humans-in-the-loop at all key decisions and AI optimization within a set model) to near-autonomous (Swan's AI agents driving a key part of the business model, supervised by humans in exceptional circumstances). With these examples in mind, let's proceed to define what constitutes an Autonomous Business Model and explore the broader implications of this phenomenon.

## 5. Defining Autonomous Business Models (ABMs)

Building on the aforementioned vignettes, we define an *Autonomous Business Model (ABM) as a business model in which agentic AI systems are the primary agents executing the firm's value creation, delivery, and capture logic, with minimal ongoing human intervention, by leveraging their capacity to sense, decide, and learn adaptively*.

This definition rests on three core mechanisms derived from the vignettes that together characterize autonomy in business models:

business model's logic. This means that an AI Factory drives how the firm interacts with customers, configures operations, and responds to changing inputs. While humans still design and supervise the system, the execution becomes AI-Factory led. In contrast, in AI-augmented models, humans retain this agency.

*Mechanism 2: Minimal human intervention*

Autonomy is not binary but measured along a continuum. ABMs reduce the frequency and necessity of human involvement in routine decisions. Human oversight becomes exception-based and only activated when the AI flags a novel scenario or when strategic goals are updated. Compared to traditional models, the operational workload for humans drops sharply.

*Mechanism 3: Adaptive decision cycles*

ABMs are not static automation systems. They are dynamic, data-driven learners which continuously self-improve in feedback loops (Iansiti and Lakhani 2020). The core loop involves sensing the environment (i.e. through data collection), deciding (i.e. via AI algorithms), learning, and improving. This continuous cycle of adaptation enables the business model to evolve in real time without manual reprogramming.

Together, these mechanisms enable ABMs to operate not just more efficiently, but *differently*. Rather than viewing the firm as a hierarchy of humans and processes, the ABM reframes it as a system governed by agentic AI loops around an AI Factory that continuously adjusts how value is created, delivered, and captured. Humans become the add-on[2].

This transformation towards ABMs is visualized in *Figure 5*. The figure depicts the progressive transformation of business models as artificial intelligence becomes increasingly central to decision-making, moving from human-driven to AI-driven logics of value creation and capture. The model distinguishes three overarching governance categories: *human-driven*, *human-AI hybrid*, and *AI-driven*.

*Augmented Business Model*, characterized by *business model enhancement*—performance improvements without fundamental shifts in agency (Rai et al. 2019).

As firms systematically integrate AI into key functions, they transition into the *Human-AI Hybrid Business Model* stage. In this phase, companies increasingly experiment with AI across functions such as personalization, forecasting, or scheduling, but humans remain the primary decision-makers and orchestrators. Governance is characterized by shared agency, with humans supervising AI outputs and retaining strategic control (Rai et al. 2019).

Subsequently, organizations transition to *Business Models with Human-in-the-Loop*. Here, AI agents begin to lead core execution tasks, with humans providing oversight, setting constraints, and intervening selectively. In this *business model delegation* transition, substantial

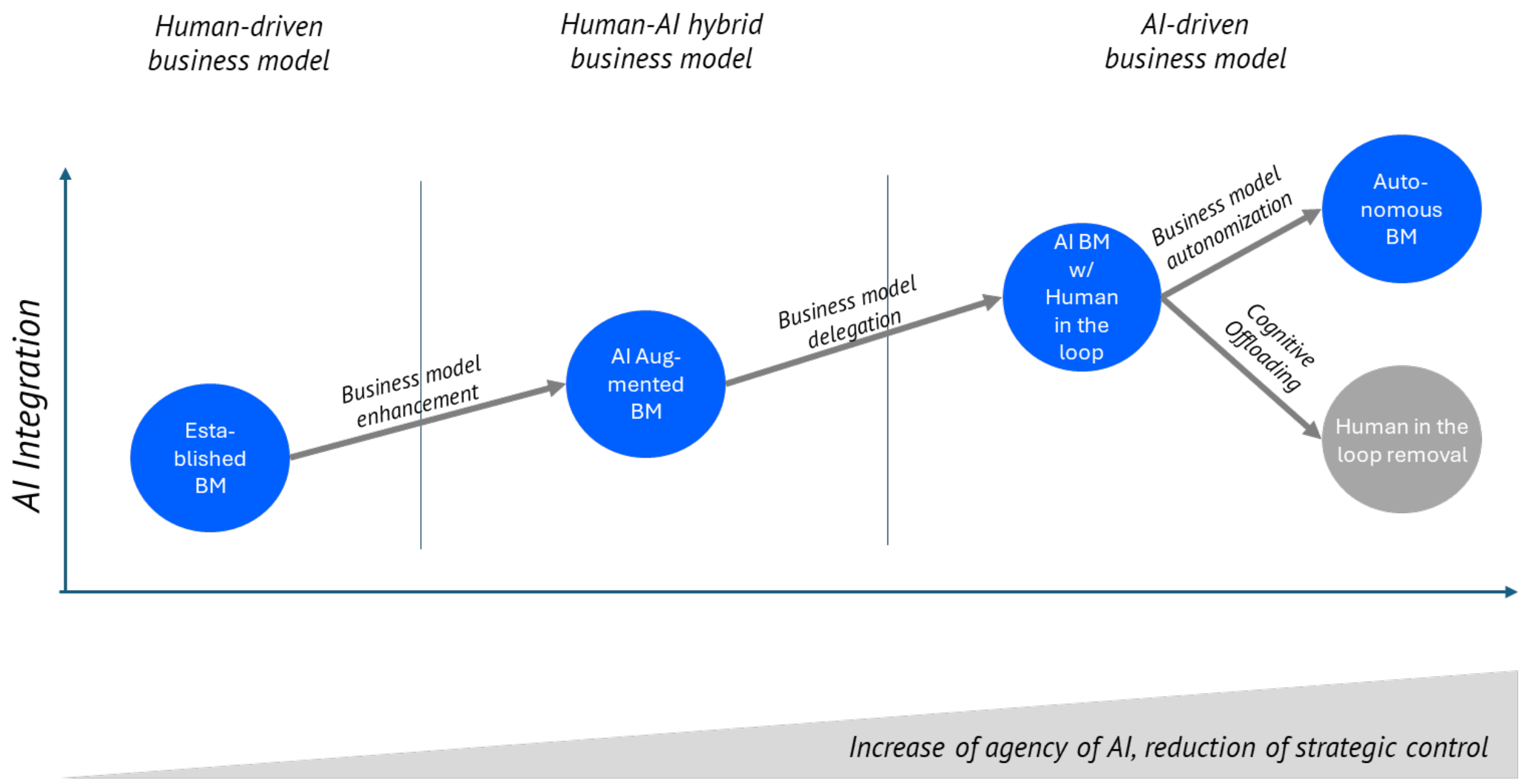

Figure 5: **The Path towards Autonomous Business Models**

The transformation begins with the *Established Business Model*, a traditional, fully human-controlled business model with no meaningful AI integration. In this *Human-Driven Business Model* stage, all strategic and operational decisions are made by humans, with AI either absent or playing a minimal, peripheral role. Firms operating in this regime rely entirely on human judgment for value creation, delivery, and capture. This is where most companies operate at the time of writing this article. As AI capabilities are embedded into isolated processes, firms transform into the *AI-* decision-making power is ceded to AI systems within bounded frameworks (Berente et al. 2021).

The next phase leads to the *AI-Driven Business Model* stage, where AI systems increasingly assume a leading role in executing, adapting, and potentially redefining the business model itself. Human involvement shifts to goal-setting, governance, and exception management. In this regime, AI agents are entrusted with core operational decisions, enabling a significant offloading of cognitive and strategic reasoning tasks (Sjödin et al. 2021). During the transition from hybrid to fully autonomous models, two critical dynamics emerge: *business model autonomization*, or the gradual transfer of operational and adaptive functions to AI agents, and *cognitive offloading* - the delegation of strategic reasoning processes such as inference, optimization, and experimentation to AI systems. Over time, this evolution leads to the *removal of the human-in-the-loop* (Gerlich 2025). These stages and their characteristics are summarized in *Table 1*.

[2] In an agent-driven model, people become the value-add rather than the workhorse, but they never cede control—human judgement and governance remain the final authority.

digitization of information flows. For instance, one could imagine an autonomous manufacturing business model (picture a "lights-out factory" that receives orders and produces goods without human labor), which is a physical product context, but heavily sensorized and digitized.

In other words, the emergence of ABMs represents a qualitative shift: the firm itself can be seen as an AI-driven system, with humans setting goals and parameters, but the "engine" of the business increasingly runs on autopilot.

Table 1: **Characteristics of Business Models Across the Autonomy Path**

| **Stage** | **Governance Logic** | **Role of AI** | **Role of Humans** | **Core Mechanisms** | **Nature of Value Creation and Capture** |
|---|---|---|---|---|---|
| **Established Business Model** | Fully human-driven | Minimal or absent | Decision-making and execution | Non-agentic AI, continuous human intervention, experiential learning | Static coordination of resources |
| **AI-Augmented Business Model** | Human-led with AI support | Decision support across selected functions | Primary decision-makers; AI enhances but does not decide | Partial agentic execution, high-frequency human oversight, tool-based learning | Enhanced efficiency within traditional value structures |
| **AI Business Model with Human-in-the-Loop** | Shared agency between humans and AI | Leading role in operations; AI proposes and acts | Supervisory role; humans intervene selectively and define objectives | Operational agentic execution, exception-based human intervention, constrained adaptive learning | Dynamic, semi-autonomous optimization of value flows |
| **Autonomous Business Model (ABM)** | AI-driven governance | Autonomous sensing, deciding, and adapting across operations | High-level strategic oversight and governance only | Full agentic execution, rare human intervention, continuous adaptive learning | Fully dynamic, real-time optimization and reconfiguration |

It is useful to delineate what ABMs are *not*. An ABM is not simply a firm that uses a lot of automation or IT. Many traditional businesses have automated processes (assembly lines, ERP software automating accounting entries, etc.), but their *business model logic* (the big decisions and coordination of how value is provided) is still driven by human judgment. It is important to stress that ABMs differ in their degree of decision-making autonomy. Likewise, ABMs are not merely "digital businesses" or "platform businesses," although being at least partially digital is a prerequisite – autonomy requires

This concept challenges some traditional assumptions in strategy —for instance, that strategic authority resides only with human decision-makers and that advantage depends on resources competitors cannot easily acquire.

But what happens when business models gain true autonomy? In what follows, we examine the implications of autonomous business models for strategy, competition, and management, exploring both the opportunities and the challenges they present.

## 6. Strategy in the Age of AI: Strategic and Competitive Implications of Autonomous Business Models

The rise of agentic AI introduces a fundamental reordering of competitive dynamics in digital markets. As firms evolve from human-driven models to Autonomous Business Models (ABMs), strategy itself begins to shift from an executive-centered process to an AI-embedded one. In this section, we examine how ABMs reshape competitive advantage, reconfigure organizational capabilities, and give rise to new forms of rivalry —what we term *synthetic competition*.

*Agentic AI as a Strategic Asset*

Agentic AI differs from earlier forms of automation and even from standard generative AI in one crucial way: it acts with *initiative*. Rather than responding passively to user prompts, agentic AI agents manage workflows, initiate actions, and optimize toward defined goals (Bommasani et al. 2023). This makes them uniquely suited to become the core drivers of business model execution.

In an ABM, agentic AI becomes a strategic asset not because it reduces costs, but because it enables *new types of value creation and adaptation*. These systems can autonomously explore product configurations, test pricing strategies, adjust supply chains in real time, and dynamically respond to customer preferences (Sjödin et al. 2021). As we argued earlier, the core mechanisms of ABMs—agentic execution, minimal human intervention, and adaptive learning—together produce a form of operational and strategic agility that exceeds human coordination capacity.

For startups like getswan.ai, this agility is native. The AI system functions as the business: identifying leads, personalizing outreach, and scheduling meetings without human involvement. For incumbents like Ryanair, reaching this level of autonomy requires layering agentic AI on top of decades of operational systems, but the principle remains the same. The strategic question becomes less about whether a firm uses AI and more about *how much agency* it grants to these systems.

*Data and Feedback Loops as Moats*

In traditional strategy, firms protect competitive advantage through barriers to imitation: proprietary technology, economies of scale, or brand equity. In the ABM context, two moats take center stage: *data loops* and *learning feedback systems*.

ABMs thrive on data. Every customer interaction, system response, and market fluctuation becomes fuel for improvement. Firms that operate ABMs gain not only efficiency, but *compounding intelligence*. As the system learns, it becomes better at achieving the firm's goals, without additional human labor. This dynamic resembles the logic of network effects (Parker, Van Alstyne, & Choudary 2016) but applied to learning: the more the system operates, the more valuable it becomes. Iansiti and Lakhani (2020) described this as the "AI factory," where data, models, predictions, and outcomes form a self-reinforcing loop.

This feedback loop becomes a strategic moat in at least two ways. First, firms that accumulate more operational data will have better-performing ABMs. Second, because agentic AI systems adapt based on their specific environment, their performance is not easily transferable. A rival firm cannot simply copy an ABM; it would also need access to equivalent data flows and operational context. These dynamics suggest a new kind of path dependency in competitive advantage: once an ABM is in place and learning, it becomes increasingly difficult to catch.

*Algorithmic Execution and Strategic Precision*

One of the implications to be explored is how they shift the locus of strategic advantage from ideation to *execution*. Traditionally, firms competed by devising superior strategies and implementing them effectively. With ABMs, execution itself becomes a differentiator. An autonomous system that can test 100 micro-pricing variations per hour or run millions of marketing permutations is not simply following a strategy; it is *strategizing through execution*.

This precision offers a level of granularity that human-managed systems cannot match. In the getswan.ai case, the AI agent adapts outreach language, response timing, and engagement patterns in real time. Such micro-optimization is impractical for human teams. In the hypothetical Ryanair ABM, an AI factory might continuously adjust route scheduling, aircraft maintenance, and ancillary service pricing based on current demand and competitive pressure. The value of the business model no longer comes from its static design but from its *dynamic responsiveness*.

Importantly, algorithmic execution also alters the nature of risk. Human strategists often face delays between decision and feedback, and ABMs shorten this loop. However, they may also propagate errors rapidly if not properly governed. Hence, firms must pair autonomy with robust oversight mechanisms. They need to develop new AI governance, auditability, and ethics initiatives and capabilities (Rai 2020).

*Ecosystem Control and Competitive Scope*

Firms building ABMs do not operate in isolation. As Jacobides, Cennamo, and Gawer (2018) argue, advantage in digital markets often derives from *ecosystem positioning*. This logic only deepens in the age of ABMs. Agentic AI systems depend on access to cloud infrastructure, real-time data streams, external APIs, and third-party platforms. Consequently, the firm that *orchestrates* or *controls* the ecosystem infrastructure can exert disproportionate influence.

This creates an asymmetry: platform leaders (e.g., Amazon AWS, Microsoft Azure, and Google Cloud) can provide the tools for ABMs while also deploying their own. Moreover, firms that control distribution points (e.g., app stores, marketplaces, ad networks) may restrict or enable the scaling of ABMs by others. This reintroduces concerns about dependency and enclosure, especially for startups. The strategic imperative becomes one of alignment: ABM builders must either develop their own ecosystems or secure privileged positions within existing ones.

Furthermore, ecosystems facilitate indirect learning. For example, an ABM that interacts with many partners and users can improve through its own proprietary data as well as through *ecosystem-level data flows*. This strengthens the data-feedback moat while broadening the ABM's capacity to adapt. In effect, an ecosystem-centric ABM competes not as a firm, but as a *distributed intelligence system*.

*Synthetic Competition and Algorithmic Rivalry*

The rise of ABMs also signals a transformation in the nature of market competition. As agentic AI systems gain autonomy, we foresee the emergence of *synthetic competition*: a regime in which AI-run business models compete directly with each other, often in real time and at machine speed.

This dynamic is already visible in algorithmic trading, where bots engage in high-frequency interactions with limited human oversight. In an ABM world, similar dynamics may unfold across domains like pricing, supply chain negotiation, customer acquisition, and even innovation. ABMs could detect a competitor's pricing move and counter it algorithmically within seconds or optimize customer targeting based on live feedback from rival offerings.

This raises two important implications. First, traditional notions of strategic positioning may lose traction. If every ABM can instantly observe and respond to competitive signals, durable advantage becomes harder to sustain. Second, the competitive advantage may hinge less on *strategy formulation* and more on *algorithm design and training*. Firms will compete through the intelligence of their agentic systems - through the amplitude of data, the quality of optimization routines, and the speed of adaptation.

Synthetic competition also poses regulatory and ethical challenges. If ABMs collude algorithmically (even unintentionally), or engage in exploitative optimization (e.g., price discrimination), the question of responsibility becomes complex. Existing frameworks for antitrust, liability, and consumer protection may need to be revised to account for algorithmic agency (Gal & Elkin-Koren 2017; Jacobides et al. 2021b).

*Redefining Strategic Capabilities and Leadership*

ABMs demand new organizational capabilities. Traditional strategic planning, emphasized vision, analysis, and top-down alignment. In ABM-driven firms, the emphasis shifts toward *AI stewardship*: the ability to specify goals, design guardrails, monitor AI performance, and intervene when needed (Iansiti & Lakhani, 2020).

This reframes leadership as meta-level governance. Executives must manage *how* decisions are made, not just *what* decisions are made. Strategic foresight becomes a matter of configuring agentic systems to explore scenarios, simulate outcomes, and self-adjust. As such, firms will need hybrid capabilities across data science, organizational design, ethics, and systems engineering.

One consequence is a shift in the profile of strategic talent. While creativity and judgment remain essential, the competitive edge may go to leaders who can conceptualize business models *as algorithms*. This includes understanding feedback loops, model architectures, and the limitations of predictive learning. In this view, the strategist is not just a visionary but an *AI architect*.

ABMs also challenge the role of organizational routines. In traditional settings, routines create stability. In ABMs, routines are codified in models that can rewrite themselves. The firm becomes more fluid but also more opaque. Hence, governance systems must evolve in parallel, monitoring performance, detecting drift, and ensuring alignment with broader objectives and values (Zeng, Guo, & Zhang 2021).

*Managerial Implications*

For managers, Figure 6 offers a strategic map for operating and competing with Autonomous Business Models (ABMs) in the age of agentic AI. At the center is the ABM itself, an intelligent adaptive system capable of executing business logic, learning from data, and interacting across networks without constant human control. Surrounding this core are three reinforcing loops that act as modern moats. Precision execution enables rapid and context sensitive responses to market signals. Compounded ABM intelligence continuously improves performance by learning from operational data. Ecosystem control allows firms to influence interdependent systems of customers, partners, and platforms.

These loops amplify one another. Precision execution generates higher quality data, which improves learning and adaptation. Compounded learning enhances the firm's ability to coordinate with external actors, and stronger ecosystem control feeds back into more targeted and effective execution. Success in this environment depends both on building these loops and on actively managing their alignment and interdependence.

introduces model competition, where AI-driven business models outperform human-driven incumbents, such as an AI-native fintech streamlining loan decisions against a traditional bank. At a broader level, ecosystem competition pits entire constellations of interdependent ABMs against one another.

For example, Amazon's retail logistics ecosystem competes with a decentralized commerce stack built on Shopify, Stripe, and OpenAI. These forms of competition converge into what can be called synthetic competition - a regime in which autonomous agents compete, learn, and optimize continuously, often beyond direct human awareness or intervention.

Leading in this environment requires a new kind of AI leadership. Rather than manually managing operations, leaders must design, integrate, and protect the systems that manage themselves. Design involves shaping the architecture of the ABM and improving its internal feedback loops. Integration focuses on connecting disparate ABMs across workflows and customer interfaces, ensuring that agentic systems interact coherently. Protection entails maintaining data quality, reliability, and security, while detecting drifts and aligning agent behavior with the firm's strategic intent and values.

At Swan AI, this leadership manifests in how founders shape the learning objectives of their

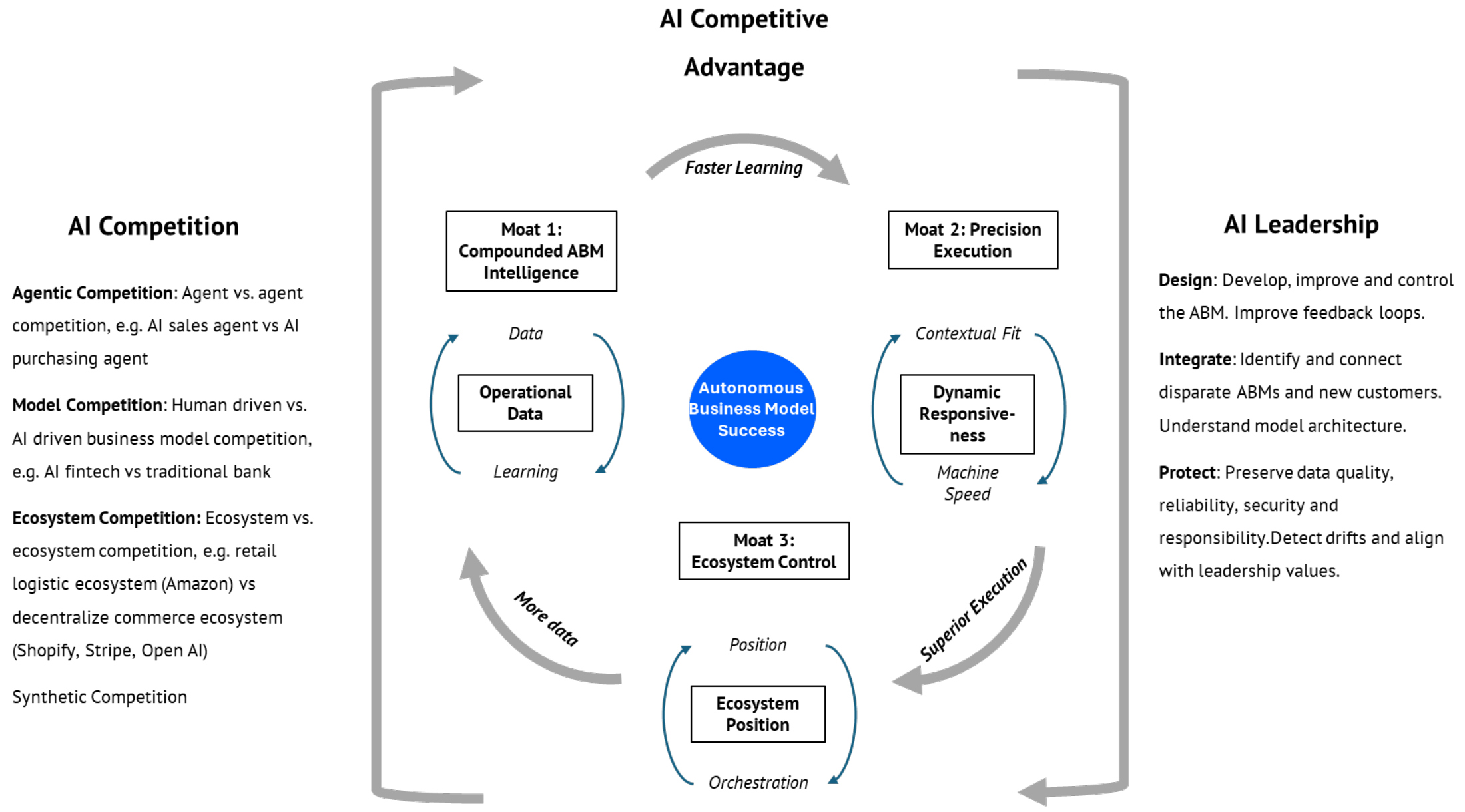

Figure 6: **Strategic Loops for Competing with Autonomous Business Models**

As agentic systems scale, firms will increasingly compete through the intelligence of their ABMs. This marks a shift toward agentic competition, where AI agents directly interact with and counter other agents, as when an AI sales agent engages with an AI purchasing agent. It also sales agents, only intervening when the system deviates from brand tone or ethical expectations. For a future-facing Ryanair, it might involve continuously refining the coordination between pricing, scheduling, and aircraft operations, while ensuring that the system remains compliant with regulatory requirements and customer expectations. In both cases, strategic advantage no longer flows from human orchestration alone, but also from the ability to lead autonomous systems that think and act with increasing independence.

## 7. Future Research Directions

Several research questions emerge from the rise of ABMs. First, how will ABMs impact workforce dynamics? A compelling example arises from getswan.ai: in the future, could an ABM decide to hire human agents because it learns that interacting with other AI agents (e.g., on the buyer's side) leads to diminishing returns in engagement? This would suggest a reversal in automation logic, where human authenticity becomes a strategic asset in synthetic markets. This is an idea supported by recent research by Xu et al. (2025), who created a simulated software development company run entirely by AI agents. In this experiment, the researchers found that AI agents failed to understand the implications and goals embedded in social conversations and they lacked the common sense and background knowledge to infer implicit assumptions. It remains to be seen of course whether agentic AI will be able to overcome these barriers.

Second, ABMs may reshape the theoretical foundations of multiple management disciplines. For instance, in the field of Top Management Teams (TMT), where studies emphasize team diversity, decision-making styles, and leadership behaviors (Hambrick & Mason 1984), ABMs may require a shift in focus toward meta-decision architectures and AI oversight structures. In Organizational Behavior, the study of motivation, engagement, and communication (e.g., Hackman & Oldham 1976) may need to adapt to environments where the majority of "organizational action" is performed by non-human agents. In International Business, where location, governance, and coordination are key themes (e.g., Johanson & Vahlne 1977), new research might explore how ABMs localize strategies or navigate cross-border regulation without human intermediaries.

These disciplinary shifts suggest the need for an integrative research agenda. How do agentic systems coordinate across boundaries, resolve ambiguity, or exert political influence? What theories of agency, control, and legitimacy apply in AI-led firms?

As business models evolve from human-led to AI-driven systems, distinct theoretical and managerial challenges arise at each stage. The table below proposes a future research agenda organized by the five stages in the progression toward autonomy. Each question reflects a dominant tension or inflection point faced by organizations operating in that mode.

Table 2: **Characteristics of Business Models Across the Autonomy Path**

| Stage | Core Research Question |
|---|---|
| ***Established Business Model*** | What foundational capabilities and data infrastructures must firms build to prepare for an eventual shift toward AI integration, and how do they recognize when to begin the transition? |
| ***AI-Augmented Business Model*** | How does the integration of AI into discrete processes reshape organizational routines, decision-making power, and perceptions of human value within the firm? |
| ***AI Business Model with Human-in-the-Loop*** | What governance models are most effective when AI systems begin to lead operational execution, and how should humans intervene without stifling system learning or adaptation? |
| ***Autonomous Business Model*** | When AI agents become the primary executors of the business model, what new forms of strategic differentiation and competitive advantage emerge, and what are the limits of autonomy? |
| ***Human-in-the-Loop Removal*** | What ethical, legal, and social boundaries must be drawn around the full removal of humans from decision loops, and under what conditions should human agency be reintroduced? |

While the stage-based agenda outlines how strategic and governance challenges evolve as firms progress toward autonomy, the next table organizes the research terrain through the lens of the three mechanisms that define ABMs: agentic execution, minimal human intervention, and adaptive learning. This structure allows for a cross-domain synthesis that aligns core management disciplines with the systemic properties of autonomous business models. It highlights how foundational theories in strategy, organization, and ethics must evolve in response to the algorithmic logics that now drive value creation, delivery, and capture.

Table 3: **Characteristics of Business Models Across Mechanisms of ABMs**

| **Domain** | **Agentic Execution** | **Minimal Human Intervention** | **Adaptive Learning** | **Seminal References** |
|---|---|---|---|---|
| | *AI performs model execution tasks* | *Humans supervise, AI acts* | *System improves decisions continuously* | |
| ***Strategic Management*** | How does agentic execution alter the sources of strategic advantage? | What are the implications of minimal human intervention for strategy formulation? | How do adaptive learning cycles in ABMs affect strategic agility and resilience? | Porter (1996); Teece (2010) |
| ***Top Management Teams (TMT)*** | How should TMT research adapt when AI systems replace middle management execution? | What oversight structures are needed for human executives in ABMs? | How do leadership roles evolve in organizations governed by adaptive AI systems? | Hambrick & Mason (1984); Carpenter et al. (2004) |
| ***Organizational Behavior (OB)*** | How does agentic AI change traditional models of motivation and job design? | What new forms of human-AI trust emerge under minimal intervention regimes? | How do employees experience work when feedback and learning are algorithmically managed? | Hackman & Oldham (1976); Organ (1988) |
| ***International Business (IB)*** | How do ABMs autonomously localize decision-making in global contexts? | What happens to international coordination when agents adapt independently? | How do regulatory differences affect agentic learning across borders? | Johanson & Vahlne (1977); Rugman & Verbeke (2001) |
| ***Human Resource Management (HRM)*** | What capabilities are needed to design and supervise agentic AI teams? | How does HR strategy shift under minimal human operational involvement? | How do organizations manage learning and development for hybrid human-AI workforces? | Wright & McMahan (1992); Lepak & Snell (1999) |
| ***Entrepreneurship*** | What design principles enable startups to build ABMs from inception? | How do entrepreneurial teams govern AI agents with increasing autonomy? | What are the funding and scaling implications of agentic business models? | Shane & Venkataraman (2000); Aldrich & Ruef (2006) |
| ***Technology and Innovation Management*** | How do innovation processes evolve when experimentation is AI-led? | What is the role of human creativity in agentic adaptive cycles? | How does innovation diffusion change under decentralized AI governance? | Tushman & Anderson (1986); Rosenberg (1990) |
| ***Business Ethics and Corporate Governance*** | What ethical principles guide agentic decision-making in business? | How can firms ensure accountability when actions are autonomously chosen? | What governance models best align adaptive AI with stakeholder interests? | Donaldson & Preston (1995); Aguilera & Jackson (2003) |

## 8. Conclusion

We are entering a new era in which agentic AI transforms not only what firms do, but how they compete, adapt, and lead. Autonomous Business Models emerge as AI agents assume responsibility for core functions of value creation, delivery, and capture. In this context, strategy ceases to be a static plan. It becomes a continuous learning system, shaped by algorithmic execution, adaptive feedback loops, and machine-speed responsiveness.

This shift introduces more than operational efficiency. It brings about synthetic competition, where intelligent systems compete, learn, and evolve in real time. It also calls for a new form of AI leadership - one that can guide the design, integration, and protection of autonomous systems aligned with organizational goals and values.

The rise of agentic AI is more than a technological disruption. It marks a structural transformation in how business models and strategy are conceived and implemented, and in how they evolve. With firms beginning to think and act through agentic systems, the fundamental challenge lies in how we design and lead organizations that operate faster and learn better than we do?

As the implications of agentic AI extend across strategy, governance, ethics, and organizational theory, we are left with a final question: Is it time to open a new cross-disciplinary conversation on agentic AI at the very core of management research, precisely the ambition this special issue set out to pursue?

## References

**Aguilera** RV, Jackson G (2003) The cross national diversity of corporate governance: Dimensions and determinants. *Acad. Manag. Rev. 28*(3): 447–465.

**Aldrich** H E, Ruef M (2006) *Organizations evolving* (2nd ed.). SAGE Publications.

**Amazon Web Services** (2019) *Ryanair improves customer support using Amazon SageMaker* [Case study]. Amazon Web Services. https://aws.amazon.com/solutions/case-studies/ryanair/.

**Amit** R, Zott C (2001) Value creation in e-business. *Strateg. Manag. J. 22*(6–7): 493–520.

**Aylak** BL (2025) SustAI-SCM: Intelligent supply chain process automation with agentic AI for sustainability and cost efficiency. *Sustainability*. *17*(6): 2453.

**Bodonhelyi** A, Bozkir E, Yang S, Kasneci E, Kasneci G (2024, January) User intent recognition and satisfaction with large language models: A user study with ChatGPT [Preprint]. *arXiv*. https://doi.org/10.48550/arXiv.2402.02136

**Bommasani** R, Hudson DA, Adeli E, Altman R, Arora S, von Arx S, ... Liang P (2023) On the opportunities and risks of foundation models. *Commun. ACM. 66*(3): 78–91.

**Carpenter** MA, Geletkanycz MA, Sanders WG (2004) Upper echelons research revisited: Antecedents, elements, and consequences of top management team composition. *J. Manag. 30*(6): 749–778.

**Casadesus-Masanell** R, Ricart JE (2010) From strategy to business models and onto tactics. *Long Range Plann*. *43*(2–3): 195–215.

**Donaldson** T, Preston LE (1995) The stakeholder theory of the corporation: Concepts, evidence, and implications. *Acad. Manag. Rev. 20*(1): 65–91.

**Deslandes** N (2025, March 20) WestJet and Ryanair CEOs: How emergencies are forcing airlines to modernise. *TechInformed*. https://techinformed.com/technical-emergencies-forced-airlines-to-digitize/.

**Gal** MS, Elkin-Koren N (2017) Algorithmic consumers. *Harv. J.L.& Tech*. *30*(2): 309–353.

**Gerlich** M (2025) AI tools in society: Impacts on cognitive offloading and the future of critical thinking. *Societies*. *15*(6), Article 10006. https://doi.org/10.3390/soc15010006

**Gregory** RW, Henfridsson O, Kaganer E, Kyriakou H (2022) Data network effects: Key conditions, shared data, and the data value duality. *Acad. Manag. Rev*. 47(1).

**Hackman** JR, Oldham GR (1976) Motivation through the design of work: Test of a theory. *Organ. Behav. Hum. Perform*. *16*(2): 250–279.

**Hambrick** DC, Mason PA (1984) Upper echelons: The organization as a reflection of its top managers. *Acad. Manag. Rev. 9*(2): 193–206.

**Iansiti** M, Lakhani KR (2020). *Competing in the age of AI: Strategy and leadership when algorithms and networks run the world*. Harvard Business Review Press.

**Jacobides**, MG, Cennamo C, Gawer A (2018) Towards a theory of ecosystems. *Strateg. Manag. J. 39*(8): 2255–2276.

**Jacobides** MG, Brusoni S, Cennamo C, Velu C (2021a) The evolutionary dynamics of ecosystem control. *Strategic Management Review*. *2*(2): 33–73.

**Jacobides** MG, Brusoni S, Candelon F (2021b) The evolutionary dynamics of the artificial intelligence ecosystem. *Strategy Science. 6*(4): 412–435.

**Johanson** J, Vahlne JE (1977) The internationalization process of the firm: A model of knowledge development and increasing foreign market commitments. *J. Int. Bus.8*(1): 23–32.

**Lepak** DP, Snell SA (1999) The human resource architecture: Toward a theory of human capital allocation and development. *Acad. Manag. Rev. 24*(1): 31–48.

**Lopez** A (2025, April 2) Beyond data: How AI is both predicting and creating consumer trends. *Forbes*. https://www.forbes.com/councils/forbesbusinesscouncil/2025/04/02/beyond-data-how-ai-is-both-predicting-and-creating-consumer-trends/.

**Markides** C (2013) Business model innovation: What can the ambidexterity literature teach us? *Acad. Manag. Perspect. 27*(4): 313–323.

**Mohammed** IA, Sofia R, Radhakrishnan GV, Jha S, Al Said N (2025) The role of artificial intelligence in enhancing business efficiency and supply chain management. *J. Inf. Syst. Eng. Manag. 10*(10s). https://www.jisem-journal.com/

**Organ** DW (1988) *Organizational citizenship behavior: The good soldier syndrome*. Lexington Books.

**Parker** GG, Van Alstyne MW, Choudary SP (2016) *Platform revolution: How networked markets are transforming the economy—and how to make them work for you*. W. W. Norton & Company.

**Porter** ME (1996) What is strategy? *Harv.Bus. Rev. 74*(6): 61–78.

**Rai** A, Constantinides P, Sarker S (2019) Next-generation digital platforms: Toward human–AI hybrids. *MIS Quarterly, 43*(1), iii–ix

**Rai** A (2020) Explainable AI: From black box to glass box. *J. Acad. Mark. Sci. 48*: 137–141.

**Rosenberg** N (1990) Why do firms do basic research (with their own money)? *Res. Policy. 19*(2): 165–174.

**Rugman** AM, Verbeke A (2001) Location, competitiveness, and the multinational enterprise. *Oxford Handbook of International Business*, 150–177.

**Shah** C, White RW, Andersen R, Buscher G, Counts S, Das SS, Montazer A, Manivannan S, Neville J, Ni X, Rangan N, Safavi T, Suri S, Wan M, Wang L, Yang L (2024, May 10) Using large language models to generate, validate, and apply user intent taxonomies. *arXiv.* https://doi.org/10.48550/arXiv.2309.13063

**Shane** S, Venkataraman S (2000) The promise of entrepreneurship as a field of research. *Acad. Manag. Rev. 25*(1): 217–226.

**Sjödin** D, Parida V, Palmié M, Wincent J (2021) How AI capabilities enable business model innovation: Scaling AI through co-evolutionary processes and feedback loops. *J.Bus. Res.* 134: 574-587.

**Swan AI**. (n.d.). *Turning visitors into deals shouldn't feel like rocket science*. Retrieved April 30, 2025, from https://www.getswan.com/

**Teece** DJ (2010) Business models, business strategy and innovation. *Long Range Plann 43*(2–3): 172–194.

**Tushman** ML, Anderson P. (1986) Technological discontinuities and organizational environments. *Adm. Sci. Q. 31*(3): 439–465.

**Wiesinger** J, Marlow P, Vuskovic V (2024, September 2) *Agents* [White paper]. Google

**Wright** PM, McMahan GC (1992) Theoretical perspectives for strategic human resource management. *J. Manag. 18*(2): 295–320.

**Xu** FF, Song Y, Li B, Tang Y, Jain K, Bao M, Wang ZZ, Zhou X, Guo Z, Cao M, Yang M, Lu HY, Martin A, Su Z, Maben L, Mehta R, Chi W, Jang L, Xie Y, … Neubig G (2024, December 18) *Theagentcompany: Benchmarking llm agents on consequential real world tasks* [Preprint]. arXiv. https://arxiv.org/abs/2412.14161

**Zeng** Y, Guo B, Zhang D (2021) Governance of artificial intelligence: Frameworks and practices. *Philos. Technol.34*(3): 543–567.

**Zott** C, Amit R (2010) Business model design: An activity system perspective. *Long Range Plann. 43*(2–3): 216–226.